\documentclass[runningheads]{llncs}
\pdfoutput=1

\usepackage{graphicx}

\usepackage{times}
\usepackage{soul}
\usepackage{url}
\usepackage[hidelinks]{hyperref}
\usepackage[utf8]{inputenc}
\usepackage[small]{caption}

\usepackage{amsmath}

\usepackage{booktabs}
\usepackage{algorithm}
\usepackage{algorithmic}
\usepackage{multicol}
\usepackage{colortbl}
\usepackage[switch]{lineno}

\usepackage[utf8]{inputenc}
\usepackage{niravstyle}
\usepackage{url}

\newcommand{\edit}[1]{\textcolor{black}{#1}}
\newcommand{\editst}[1]{}

\usepackage{xspace}
\newcommand{\method}{\fsl{Svoie}\xspace}
\newcommand{\baseline}{\fsl{Stable-SVO}\xspace}
\newcommand{\timesteps}{\np{1000}\xspace}
\newcommand{\numagents}{\np{300}\xspace}

\usepackage{multirow}
\usepackage{subcaption}
\usepackage{cleveref}
\usepackage{graphicx}
\usepackage{booktabs}

\usepackage{listings}

\usepackage{tikz}
\usetikzlibrary{fit}
\usepackage{pgfplots}
\pgfplotsset{compat=newest}
\usetikzlibrary{pgfplots.statistics}
\usetikzlibrary{positioning,shapes,arrows,shadows,patterns}

\usepackage{siunitx}
\usepackage[np]{numprint}
\npthousandsep{,}
\npdecimalsign{.}

\usepackage[square,numbers]{natbib}

\newcommand{\shortcite}[1]{\cite{#1}}

\begin{document}

\title{Social Value Orientation and Integral Emotions in Multi-Agent Systems}

\author{Daniel E. Collins \orcidID{0000-0002-1075-4063} \and
Conor Houghton \orcidID{0000-0001-5017-9473}  \and
Nirav Ajmeri \orcidID{0000-0003-3627-097X}}
\authorrunning{D.~E. Collins et al.}
\institute{Department of Computer Science, University of Bristol, Bristol, UK
\email{\{daniel.collins,conor.houghton,nirav.ajmeri\}@bristol.ac.uk}
}

\pagenumbering{arabic}
\thispagestyle{plain}
\pagestyle{plain}

\maketitle

\begin{abstract}
Human social behavior is influenced by individual differences in social preferences. Social value orientation (SVO) is a measurable personality trait which indicates the relative importance an individual places on their own and on others' welfare when making decisions. SVO and other individual difference variables are strong predictors of human behavior and social outcomes. However, there are transient changes human behavior associated with emotions that are not captured by individual differences alone. Integral emotions, the emotions which arise in direct response to a decision-making scenario, have been linked to temporary shifts in decision-making preferences. 

In this work, we investigated the effects of moderating social preferences with integral emotions in multi-agent societies. We developed \method, a method for designing agents which make decisions based on established SVO policies, as well as alternative integral emotion policies in response to task outcomes. We conducted simulation experiments in a resource-sharing task environment, and compared societies of \method agent with societies of agents with fixed SVO policies. We find that societies of agents which adapt their behavior through integral emotions achieved similar collective welfare to societies of agents with fixed SVO policies, but with significantly reduced inequality between the welfare of agents with different SVO traits. We observed that by allowing agents to change their policy in response to task outcomes, agents can moderate their behavior to achieve greater social equality. 
\end{abstract}

\keywords{Individual Differences \and Social Decision-Making \and Simulation.}

\section{Introduction}
Social value orientation (SVO) is a spectrum of personality traits which describes individual differences in social preferences, in terms of the relative value an agent places on its own welfare and the welfare of others when making decisions \cite{mcclintock_social_1972,mcclintock_social_1989}. The SVO spectrum include agents who are: \fsl{altruistic} or caring only for others, \fsl{cooperative} or caring both for self and others, and \fsl{selfish} or caring only for self. SVO is measurable in humans and considered to be relatively stable over time. Further, SVO has been found to be strongly correlated with patterns of social behavior through empirical study, such as the tendency to act cooperatively or individualistically \cite{balliet_social_2009,bogaert-social-2008}. 

Seminal works from social psychology provide \edit{a} clear conceptual model of the influence of SVO on individual preferences in social interactions. A robust framework for agent simulation has been developed, the ring model \cite{liebrand_computer_1996,liebrand_ring_1988}, which defines utility functions for SVO traits that are now standard in multi-agent research. In a social dilemma, a rational agent would be expected to \edit{make} decisions which maximise the utility associated with their individual preferences. However, \editst{in isolated decision tasks,} humans \edit{are not rational agents, and will not} \editst{do not} always seek optimal outcomes that would be expected \editst{from} \edit{for} their stable characteristics. Through empirical studies, patterns of irrational decision-making in humans have been linked to transient changes in affective state, emotions and mood states, resulting from changes in immediate circumstance or environment, or the consequence of longer-term contingencies or goals which interact with the current task. Emotions may serve an important role in adaptive decision-making by motivating and guiding behavior based on observations and judgements about the current context of the decision-making environment and other within it. 

Lerner \etal \cite{lerner_emotion_2015} outline two main categories of emotion, \fsl{incidental} and \fsl{integral}. \fsl{Incidental emotions} are task-unrelated emotions which arise in response to factors which are irrelevant to the current decision scenario, but which are nevertheless present during the decision-making process. \edit{For example, a person who receives a frustrating message from a friend before an important meeting at work may be influenced by the unpleasant emotions during the meeting, even though they are task-unrelated, which could lead to impulsive decision making or unnecessary conflicts with colleagues.} \fsl{Integral emotions} are task-related emotions which arise in direct response to the current decision, and are known to have a strong influence on behavior. Integral emotions can be either \fsl{anticipated}, feelings about a  potential future event or the possible outcome of an action, or \fsl{immediate}, feelings about a recent event or the observed outcome of an action. Our interest is in the latter, for example, the immediate integral emotion of feeling satisfied after performing well on an exam, and choosing to spend time helping others with their studies.  

The ``wounded pride'' model of integral emotion \cite{zheng_influence_2017} suggests that agents may react to  unfair outcomes by feeling negative emotions, and acting spitefully, even when they know that it will result in a worse outcome for themselves on that specific task \cite{pillutla1996unfairness}. This is an example of how integral emotions can give rise to behavior that is not explained by individual differences alone. Agents which adapt their policies based on integral emotion as in the wounded pride model may fare better than agents which only act based on SVOs, since some SVO policies may perform poorly on a given task than others. In this work, we investigated whether socially beneficial effects of altering social preferences according to integral emotions could be observed by modelling integral emotions in multi-agent societies with individual differences in SVOs.

\noindent\fbf{Contributions.}
We developed \method, a method for designing agents which make decisions based on SVO and integral emotions. Our \method agents combine well-established SVO decision-making policies with a simple protocol for temporarily adopting alternative policies based on integral emotion. We define two alternative social preference based policies representing \fsl{positive} and \fsl{negative} emotions, which minimise or maximise payoff inequity respectively. These policies incorporate the wounded pride model of spiteful human decision-making, and an idealised counter model for positive integral emotion. We model integral emotion as an internal state, which changes depending on the outcomes of recent decisions, and which defines the probability that an agent will adopt an emotion based policy in their next decision.

\noindent\fbf{Findings.}
To evaluate \method, we conducted simulation experiments using a variant of the Colored Trails game \cite{gal-colored-2005-temp,grosz-influence-2004}, a resource-sharing task environment designed for studying social decision-making.  We generated societies of agents with heterogeneous SVOs, and simulated sequences of games between random pairs of agents in the society. We compare the distribution of payoffs accumulated by agents between \method and \baseline societies, and evaluate societal outcomes in terms of collective welfare, a measure of the total payoff to all agents in a society, and welfare inequality, a measure of the variation of payoff between agents.

We investigated whether \method societies would have lower welfare inequality relative to \baseline societies, by allowing agents to moderate their social preferences based on the frequency with which they are succeeding or failing to achieve their goals. We find that societies of \method agents exhibit significantly lower welfare inequality than \baseline agents in societies with more than one SVO, with a small reduction in collective welfare. 

\noindent\fbf{Organisation.}
Section~\ref{sec:preliminaries} describes preliminaries necessary to understand our contribution.
Section~\ref{sec:method} describes our method for modelling SVO and integral emotions in agents. 
Section~\ref{sec:experiment-and-results} presents our experimental setup and the results. 
Section~\ref{sec:conclusion} concludes with a discussion of future directions. 

\section{Preliminaries and Related Works}
\label{sec:preliminaries}
We now introduce the preliminaries necessary to understand our contributions.

\subsection{Social Value Orientation}
\label{sec:social-values}
The SVO model describes a continuum of orientation types, reflecting the nature of social preferences in decision-making \cite{mcclintock_social_1972,mcclintock_social_1989}. SVOs are used in agent-based simulation to define agent decision-making policies. SVO policies are typically implemented using the ring model of SVO \cite{liebrand_ring_1988}. In this model, an SVO utility functions can be defined by any point on a unit circle, where the extent of preference for reward to self and to others is mapped to the $x$ and $y$ axes respectively. For example, this spectrum includes:

\begin{description}[leftmargin=0em]
\item[Altruistic] Preference to take actions which increase the welfare of others, regardless of their own welfare.
\item[Cooperative] Preference to work with others to increases the welfare of themselves as well as others.
\item[Selfish] Preference to take actions which increase the welfare of themselves, regardless of the welfare of others.
\end{description}

These three SVOs types cover the positive quadrant of the ring model, in which SVO utility functions only consider positive preferences reward to self, other or both. The complete spectrum of SVO traits also includes negative preferences, for example, competitive agents have a preference for increasing their own reward and also reducing the reward of others. Different SVO decision making policies are well defined, and give predictable differences in performance in simulated social task environments \cite{liebrand_computer_1996}. The relative performance of SVO policies depends on the nature of the task. 

Social preferences have been investigated in the context of developing autonomous agents for applications in various real-world domains such as cyber-security \cite{moallem_social_2019}, and SVO has been used to simulate social behavior in autonomous vehicle decision-making \cite{buckman_sharing_2019,crosato_human-centric_2021,schwarting_social_2019}. Multi-agent simulation incorporating SVO has been used alongside experimental data to better understand how individual differences can influence cognition and behavior to benefit societies, e.g., through social cooperation \cite{andrighetto_cooperation_2020} and adapting to changes in environment \cite{vilone_evolutionary_2021}, and SVO has been used in the simulation of normative multi-agent systems to understand the emergence of prosocial and cooperative behavior \cite{ajmeri_fleur_2022}. Related works have looked at agent-based modelling of other individual difference variables, such as Myers-Briggs personality types \cite{braz_using_2022}.  In this work, we aim to better understand the relationship between emotion and social preferences through agent-based simulation.

\subsection{Integral Emotions}
\label{sec:integral-emotions}
Integral emotions describe task-related emotions which are directly influenced by the current decision-making process, for example, an individual may experience positive or negative integral emotions depending on whether they achieve their goal on a particular task \cite{zheng_influence_2017}. 

Seminal works from psychology shed light on the influence of integral emotions on human behavior through empirical studies using ultimatum games \cite{harsanyi_rationality_1961}. An ultimatum game between two agent, Alice and Bob, can be described as follows: Alice and Bob are in separate rooms. Alice is told that Bob has been given an amount of money, and has been asked to share some of this money with Alice. Bob can offer any portion of the money to Alice that they choose. Alice can either accept this offer, or reject it. If Alice rejects the offer, neither Alice nor Bob receive any of the money, hence Bob's offer is an ultimatum.

A key findings of early work on ultimatum games is that people often reject small amounts of money despite the fact that this results in a worse outcome for themselves --- they are rejecting ``free money''. This finding has been replicated in numerous studies \cite{yamagishi_search_2014}. This may be thought of as a calculated spiteful behavior, e.g. paying a cost in order to harm another. Emotional reactions like spite may be considered in the context of social norms, pervasive expectations of certain behaviors within societies. Spiteful actions, in which a cost is paid to punish a perceived wrongdoer, may be adaptive behaviors which encourage cooperation norms, by enforcing sanctions in the form of punishments when cooperation norms are violated \cite{Nardin-KER16-Classifying}. This could be extended to any norm related to how an individual expects that others should behave in a society, regardless of how they do. If an individual has a strong expectation for a particular norm, they may experience negative emotions when that norm is violated, and respond with spiteful actions. Behavior of this nature is common in online communities, for example, in commenting behaviors on the website Stack Overflow, \cite{aler_tubella_norm_2021}. 

The perspective of emotions as norm enforcing mechanisms is complicated by observations from ultimatum game experiments which show that spiteful behavior may arise in the absence of any perceived social injustice, in the absence of any punishable perpetrator, and that once triggered, spiteful behavior may be sustained and subsequently directed towards others arbitrarily. By altering the set-up of the ultimatum game, Straub and Murnigham \shortcite{straub1995experimental} observed that participants sometimes rejected small offers even if they did not know the total amount of money from which the offer had been made, suggesting the rejection is not motivated by a sense of social inequity between participants. Further, they found that participants were just as likely to reject small offers when they did not know that the money had been split by another participant. They hypothesized that offers of small amounts of money were rejected because they evoked feelings of wounded pride, a direct emotional response to an unsatisfactory outcome. Pillutla \etal \shortcite{pillutla1996unfairness} conducted experiments using a sequence of ultimatum games between different pairs of participants, and found that participants who spitefully rejected a small offer would be more likely to take spiteful actions in subsequent games against new participants. In ultimatum games, individuals who receive an unsatisfactory offer may still try to act in retaliation, even if they cannot cause a disadvantage to the proposer of the unfair offer, suggesting that spiteful actions are a form of emotional release, or an expressions of internalised emotions \cite{yamagishi-private-2009}. The emotion may arise due to norm violation, but the resulting action may not be a calculated effort to enforce that same norm. More recently, Criado \etal \cite{criado_human-inspired_2013} have explored role of emotions as motivators for norm compliant decision-making towards the development of autonomous agents act in accordance with human norms.

These works describe a model of wounded pride, in which undesirable task outcomes can provoke a strong negative emotional response, which is expressed through subsequent non-cooperative behavior. If an agent perceives that an outcome is unfair and unduly negative to them or contrary to an expectation of self-worth, feelings of wounded pride and anger are aroused which will influence their subsequent actions even if those actions cannot lead to a redress of the perceived wrong. In other words, when an individual experiences negative emotions in response to an unsatisfactory outcome, but cannot directly express these emotions to some perceived wrongdoer, they are nevertheless willing to retaliate by making sub-optimal decisions, which disadvantage others at some cost to themselves. This mechanism may be beneficial in protecting altruistic agents from being repeatedly taken advantage of by self motivated agents. Conversely, we can conceive of a counter mechanism to wounded pride, wherein disproportionate success may illicit positive emotions, which in turn influence an agent to temporarily relax their preferences for high payoff and promote generosity. This aligns with ideas from social psychology on behavior changes associated with positive emotion
\cite{isen_influence_2001,vastfjall_arithmetic_2016}.

A common method of monitoring integral emotions in human studies is self reporting of emotion valence, the degree of positive feeling or negative feeling at a particular moment in time. This derives from the appraisal theory of emotion \cite{zeigler-hill-appraisal-2017}, which posits that human emotions are internal phenomena, constructed through the appraisal of external events and stimuli, for example, by evaluating whether an event outcome aligns with personal goals or norm expectations. Valence has been used in autonomous agent research to define internal states related to emotions, for example, to define intrinsic rewards for guiding the behavior of reinforcement learning agents \cite{huang_computational_2021}, and as a component of comprehensive decision-making architectures based on psychological theories \cite{bosse_fatima_2014}. These related works often make use of other components of appraisal theory, such as arousal and motivation. For simplicity, we will focus on the valence of integral emotion associated with task outcomes. A similar approach has been taken previously to investigate the relationship between emotions and behavioral norms \cite{de_melo_interplay_2020}.

\subsection{Social Task Settings}
 Simulations of agent behavior in game environments can be directly compared to human decision-making data on the same or similar tasks, or used as an abstraction of complex real-world social decision-making scenarios. In stochastic games, random variations in the parameters of the games setup and the agents involved in the game can give rise to a variety of different emergent scenarios. Sequences of stochastic games of varying complexity have been used to approximate complex real world task environments for studying the influence of emotion and social factors on behavior, both in empirical human studies and agent simulation \cite{cheng_life_2011,crosato_human-centric_2021}. There is a breadth of work in which stochastic games have been used to study the relationship between SVO and social behavior \cite{balliet_social_2009}. Stochastic games have also been used to study how emotions influence behavior. Bono \etal \cite{bono_roles_2020} use a stochastic resource-allocation game to study how emotions mediate SVO preferences in human decision-making.  

Colored Trails (CT) \cite{gal-colored-2005-temp,grosz-influence-2004} is a research test-bed designed for studying social factors in decision-making. In CT, agents enter into a negotiation \cite{Kraus-AIJ97-Negotiation} and exchange resources to achieve their own individual goals. CT can be described as in terms of generic elements of the task setting:

\begin{itemize}
    \item Agents have individual goals they try to bring about.
    \item Agents have individual resources they can use to bring about their individual goals.
    \item Agents receive a reward on bringing about their goals. 
    \item Individual circumstances of agents may vary, and therefore they may require different resources to bring about their goals compared to their peers. 
    \item Agents may have insufficient resources to bring about their goals, or they may have surplus resources.
    \item Agents may negotiate to exchange resources to facilitate each other reach their goals. 
\end{itemize}

CT is a highly flexible and expressive stochastic game, with various parameters that can be modified to customise the task environment. We chose to adopt CT as an environment for evaluating our agent societies, as it benefits from a clear task setting, and the random elements in the games set-up
allow agents to encounter different unique social tasks over a sequence of games \cite{Jong2011MetastrategiesIT}.

\section{Method}
\label{sec:method}
We now detail our implementation of the CT game environment, agent decision-making policies, and agent models.  

\subsection{Simulation Environment}
\label{sec:environment}  
We implemented a simplified version of CT as a simulation environment for studying \method and \baseline agent societies. A game of CT is played between two agents, and consists of two separate rounds. At the start of each game round, a new game-board is generated: a 4$\times$4 grid of colored tiles, where each tile is randomly assigned one of four possible colors (red, blue, green, yellow). Each agent is then placed on the game-board at separate random starting positions. A random goal position is then assigned on the game-board, which is not vertically or horizontally adjacent to either agent's starting positions. At the start of each round, each agent is allocated resources --- a set of four randomly colored chips --- which agents can place to move to an adjacent position on the board where the chip colour matches the tile colour. The objective of the game is to move as close as possible to the goal position from the starting position using the allocated resources. We assume agents have access to full information about the state of the game, e.g., the game board, agent positions, goal position, and the resources of both agents. 

Once per round, the agents may negotiate and exchange some or all of their resources to help each other reach the goal. During negotiation, one agent takes the role of \fsl{Proposer} and the other takes the role of \fsl{Responder}. The \fsl{Proposer} sends a proposal to the \fsl{Responder} comprising an offer, chips they will send from their own inventory, and a request, chips they want to receive from their opponents inventory. The \fsl{Responder} can then either accept the proposal, initiating the proposed exchange, or decline the proposal, meaning there is no-exchange and both players are left with their original allocated resources. Agents can then use their resources to move as close as possible to the goal position, and receive a score, $S$, at the end of the round: 
\begin{equation}
S=n + 1.5u(1 + g) \label{eq:1}   
\end{equation}
where $n$ is the number of unused chips remaining in the agents inventory, $u$ is the number of tile-chips used to create a path and $g$ is equal to 0 or 1 depending on whether or not the agent reached the goal position respectively. This scoring function is taken from \cite{kalia-interplay-2019}, and is designed to prioritise goal achievement strategies over strategies which seek to maximise score by gathering tiles, or creating long paths to arbitrary positions. Agents swap negotiation roles between the two rounds of the game, so that each agent has one round as \fsl{Proposer} and one round as \fsl{Responder}.

By only allowing one offer and response per game round, CT becomes a more expressive form of the traditional ultimatum game discussed in Section~\ref{sec:preliminaries}. Here, the \fsl{Responder} can only choose between two possible outcomes: the ultimatum offer sent by the \fsl{Proposer}, or the no-exchange outcome determined by the randomised parameters of the game set-up. Random variations in individual circumstances and individual goals are encoded in CT through random variations in game board set up, resource allocation, starting positions and goal positions. 

\begin{figure*}[!htb]
\centering
\includegraphics[width=\textwidth]{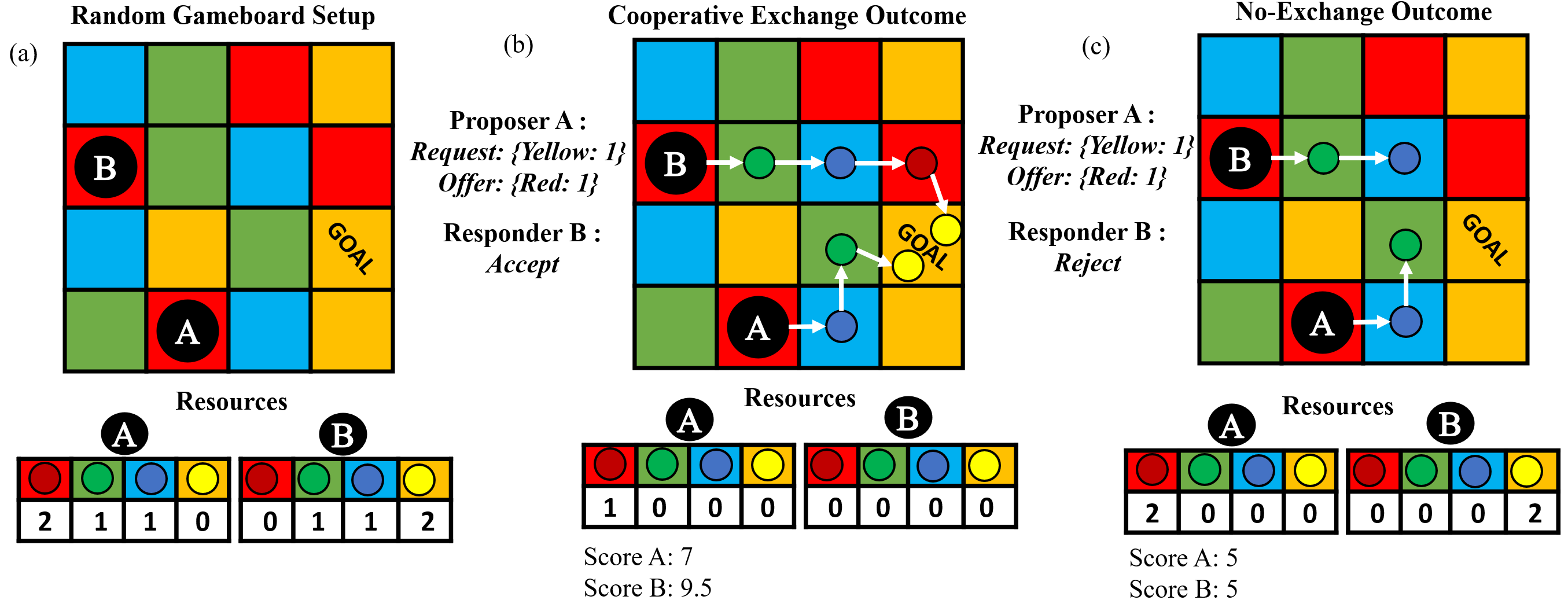}
\caption{Schematic example of one round of CT between agents A and B. (a) Random game board setup parameters are generated at the start of the game: colored tiles, agent positions, goal position and allocated resources. In CT, the resources are colored chips which agents can use to move to an adjacent tile with the same colour. In this illustrated setup, neither agent can reach the goal using their initial resources. (b) A possible game outcome is shown, in which B has agreed to A's mutually beneficial exchange proposal; A sends one red chip to B, and B sends one yellow chip to A. Agents then use their resources to reach the goal, and receive a score according to Equation~\ref{eq:1}. (c) Alternatively, in the no-exchange outcome, B chooses to reject A's proposal, and agents must move as close to the goal as they can with their initial resources. Here, this results in a lower score for both agents.}
\label{fig:1-CT-game}
\end{figure*}

Figure~\ref{fig:1-CT-game} shows a schematic example of one possible CT set-up, demonstrating how agents can cooperate to achieve a greater reward.

\subsection{Utility Functions for Social Preferences}
\label{sec:utility-functions}
To design agent decision-making protocol for the CT environment, social preferences and possible actions were mapped to quantitative utility functions. In each case, the agent perceives their environment, and uses the available information to select an action. An action is selected it is expected to maximise the utility associated with the agents social preferences, a function of the game scores expected to result from an action, calculated using the scoring function in Equation~\ref{eq:1}. The way in which an agent uses the utility function depends on whether it is acting as a \fsl{Proposer} or \fsl{Responder}.

Let, $x$, be an arbitrary exchange outcome, e.g., the resources that each agent possesses after the an exchange. If we assume that an agent will always use their chips optimally to achieve the highest possible score, each exchange outcome $x$ maps directly to a pair of scores $S-P(x)$ and $S-R(x)$ for the \fsl{Proposer} and \fsl{Responder} respectively for a given game set-up. We can therefore define our utility functions in terms of $x$. 

An agent acting as \fsl{Proposer} uses a chosen utility function as a ranking criteria to select a proposal. The \fsl{Proposer} calculates the utility associated with each exchange outcome, $x$, from the set of all possible exchange-outcomes, $X$, then selects the outcome with the greatest utility, and sends the corresponding proposal that would result in that outcome if accepted by the \fsl{Responder}. A \fsl{Responder} will accept a proposal only if it maximises a utility-based acceptance criteria relative to the no-trade outcome $\bar{x}$, the random set of resources possessed by each agent if no-exchange takes place. Here, the expected score for the no-trade outcome, $\bar{x}$, can be denoted $S-P(\bar{x})$ and $S-R(\bar{x})$. A proposal is only accepted if the utility of the proposed exchange is greater than the utility of the no-exchange outcome for the \fsl{Responder}.

Utility functions for socially oriented decision-making protocols are outlined for CT \cite{gal-agent-2010, gal-modeling-2007-temp} based on different social preferences. We adapted these utility functions to describe agent protocols for out implementation of CT:
\begin{description}[leftmargin=0em]
\item [Individual Benefit] the utility is the proposer score
    \begin{equation}
         U-r(x) = \mathit{S-R(x)}
         \label{eqn:individual-p}
    \end{equation}
    or the responder score.
       \begin{equation}
         U-p(x) = \mathit{S-P(x)}
         \label{eqn:individual-r}
    \end{equation}
    
\item [Aggregate Benefit] the utility is the cooperative score, the sum of the proposer and responder scores. 
    \begin{equation}
         U-c(x) = S-P(x) + S-R(x)
    \end{equation}
\item [Outcome Fairness (Advantage of Outcome)] the utility is the advantage achieved by the responder.
    \begin{equation}
         U-a(x) = S-R(x) - S-P(x) 
    \end{equation}
\item [Trade Fairness (Advantage of Trade)] the utility is the advantage achieved by the responder, relative to rejection.
    \begin{equation}
        U-f(x, \bar{x}) = \mathit{(S-R(x) - S-R(\bar{x})) - (S-P(x) - S-P(\bar{x}))}
    \end{equation}
\end{description}

It is important to note that these functions are written from the perspective of the \fsl{Responder} so that they are positive when the action benefits the \fsl{Responder}. When used by the \fsl{Proposer}, the subscripts \textit{P} and \textit{R} are swapped.  

\subsection{Agent Decision-Making Policies}
\label{ssec:agent-design}
In this section, we adapt the social preference based utility functions outlined in Section~\ref{sec:utility-functions} to construct decision-making policies corresponding with altruistic, selfish and cooperative SVOs, and positive and negative integral emotions. We use these policies to develop baseline \baseline agents, which always make decisions according to a fixed SVO based policy, and \method agents, which act according to an SVO based policy by default, but may temporarily adopt an integral emotion based policy in response to game outcomes in CT.

\subsubsection{SVO policies.}
\label{ssec:method:svo-policies}
Baseline \baseline agents were created such that each agent has one of three possible SVO traits: selfish, altruistic or cooperative. Each SVO describes a fixed decision-making policy with a utility function reflecting social outcome preferences.  

\begin{description}[leftmargin=0em]
    \item [Selfish] A selfish agent takes actions which maximise their own payoff.
        \begin{itemize}
            \item Proposal Ranking Criteria:     
                \begin{equation}
                \textit{maximise}\,U-p(x)
                \end{equation}
            \item Response Acceptance Criteria: 
                \begin{equation}
                \textit{accept trade if and only if}:\,U-p(x) > U-p(\bar{x})
                \end{equation}
        \end{itemize}
    \item [Cooperative] A cooperative agent takes actions which maximise mutual payoff.
        \begin{itemize}    
            \item Proposal Ranking Criteria:
                \begin{equation}
                \textit{maximise}\,U-c(x)
                \end{equation}
            \item Response Acceptance Criteria: 
                \begin{equation}
                \textit{accept trade if and only if}:\,U-c(x) > U-c(\bar{x})
                \end{equation}
        \end{itemize}
    \item [Altruistic] An altruistic agent takes actions which maximise payoff to others.
        \begin{itemize}
            \item Proposal Ranking Criteria: 
                \begin{equation}
                \textit{maximise}\,U-r(x)
                \end{equation}
            \item Response Acceptance Criteria: 
                \begin{equation}
                \textit{accept trade if and only if}:\,U-r(x) > U-r(\bar{x})
                \end{equation}
        \end{itemize}
\end{description} 

\subsubsection{Integral Emotion Policies.}
\label{ssec:method:ie-policies}
We devise two integral emotions policies to capture temporary changes in social preferences resulting from positive or negative integral emotions. Here, the integral emotion policies describe social outcome preferences that are not captured in  \baseline policies. The negative emotion policy, \fsl{competitive equity aversion}, is one which is expected to result in achieving a higher score with the largest margin of difference between the agent and its opponent (``Advantage of Outcome'') or ``unfair'' proposal). Conversely, the positive emotion policy, \fsl{inequity aversion}, is one which will minimise the margin of difference between the resulting scores. These are distinct from SVO policies as they do not consider game score maximisation.

\begin{description}[leftmargin=0em]
    \item [Positive Integral Emotion (Inequity Aversion).] An agent with positive integral emotion valence takes ``fair'' actions which minimise the difference in payoff between themselves and others.
        \begin{itemize}
        \item Proposal Ranking Criteria:
        \begin{equation}
        \textit{minimise}\,1/(1+|U-a(x)|) 
        \end{equation}
        \item Response Acceptance Criteria: 
        \begin{equation}
        \textit{accept trade if and only if}:\, 
        U-f(x, \bar{x}) < 0
        \end{equation}
        \end{itemize}

    \item [Negative Integral Emotion (Competitive Equity Aversion).] An agent with negative integral emotion valence takes ``unfair'' actions which maximise the difference in payoff between themselves and others, and for which the payoff to themselves is greater than that to others. 
        \begin{itemize}    
        \item Proposal Ranking Criteria:
        \begin{equation}
        \textit{maximise}:\,1/(1+|U-a(x)|) 
        \end{equation}
    \item Response Acceptance Criteria: 
        \begin{equation}
        \textit{accept trade if and only if}:\,U-f(x, \bar{x}) > 0
        \end{equation}
        \end{itemize}
\end{description}

\subsubsection{Internal Emotion State for \method.}
\label{sss:emo}
We adopted standard decision-making protocols for altruistic, cooperative, and selfish SVOs to form  baseline \baseline agents, where agents always make decisions which align with their SVO. We then introduced an integral emotion component to the \baseline agents to produce a \method agent --- an agent that has an SVO, as well as positive and negative integral emotion policies. We designed \method agents so that positive integral emotion would be associated with reaching the goal in a round of CT, and negative emotion with not reaching the goal. To encode integral emotion in \method, we define an internal state $E \in \{-1, -0.5, 0, 0.5, 1\}$ representing the current valence of the agent, e.g. the positiveness or negativeness of the agents integral emotion. This is an internal state that is updated based on goal achievement at the end of each game round. For simplicity, we allow $E$ to take one of five discrete states between --1 and 1, however a higher granularity or continuous implementation could be used. 

In the CT game, goal achievement results in a step increase in $E$ and conversely, goal non-achievement results in a step decrease. We use $E$ to define the probability that an agent selects an integral emotion based policy. $E = 0$ represents a neutral emotion state, in which the agent always defaults to its baseline SVO decision-making policy. When $E = 0.5$ or $E = -0.5$, the agent will have a 50\% chance of selecting the positive or negative emotion policy respectively,  and when  $E = 1$ or $E = -1$, the agent will always select the associated emotion policy. In this way, agents can exhibit varying degrees of emotion based behavior over many repeat interactions depending on how frequently their decision-making policy causes them to achieve or miss their goals. The state $E$ is designed to reflect the ``appraisal theory'' of emotion \cite{zeigler-hill-appraisal-2017}, which posits that human emotions are internal phenomena, constructed through the appraisal of external events and stimuli, for example, by evaluating whether an event outcome aligns with personal goals or expectations.

\section{Experiments and Results}
\label{sec:experiment-and-results}
We conducted simulation experiments using CT (Section~\ref{sec:environment}) as a task environment. We repeat our experiments using four different agent societies, which we define based on the proportions of agents with different SVO trait:
\begin{description}[leftmargin=0em]
    \item[\fsl{altr-coop}] Agent society with equal number of altruistic and cooperative agents
    \item[\fsl{altr-self}] Agent society with equal number of altruistic and selfish agents
    \item[\fsl{coop-self}] Agent society with equal number of cooperative and selfish agents
    \item[\fsl{mixed}] Agent society with number of altruistic, cooperative and selfish agents
\end{description}

Each simulation is run over \timesteps time steps. At each time step, each agent in the society is paired with another agent at random, and each pair of agents plays two rounds of CT and receives a score. We compare simulations of \method agent societies to simulations of \baseline societies.

\begin{description}[leftmargin=0em]
    \item[\baseline] Agents follow fixed decision-making rules associated with their SVO.
    \item[\method] Agents act the same as \baseline initially, and have an SVO trait, but may deviate from their stable SVO trait based on game outcomes. 
\end{description}

We define metrics and hypotheses in Section~\ref{sec:metrics-hypotheses} for evaluating whether the integral emotion mechanism introduced in \method has a beneficial effect on societal outcomes at the end of the simulations.

\subsection{Evaluation Metrics and Hypotheses}
\label{sec:metrics-hypotheses}
We define and compute \fsl{Individual Welfare}, \fsl{Collective Welfare} and \fsl{Welfare Inequality} for evaluating simulated \method and \baseline agent societies.

\begin{description}[leftmargin=0em]
    \item[Welfare] measures the success of agents in maximising their score. We use the mean score achieved for an individual agent, or for a sample of agents, to evaluate welfare.
    \item[Inequality] measures inequality of outcomes between members of an agent society. 
    We assess inequalities over distributions using the Coefficient of Variation (CoV) measure \cite{maio-income-2007}. Whereas Gini Coefficient is used in other research to measure inequality, we select CoV for its simplicity, and because the distributions of individual measures are observed to be approximately normal in preliminary runs.
\end{description}

\begin{enumerate}
\item \fsl{\textbf{Individual Welfare}} The mean score an individual agent achieves over all time steps in a simulation run.
\item \fsl{\textbf{Collective Welfare}} The mean score over a sample of agents.
\item \fsl{\textbf{Welfare Inequality}} The CoV of the distribution of individual welfare of agents in a sample. The magnitude of this measure is smaller for more equal societies.
\end{enumerate}

We evaluate two hypotheses corresponding to the evaluation metrics for simulated agent societies.
\begin{description}[leftmargin=0em]
    \item[H1] \method gives greater collective welfare than \baseline over all agents in a society.
    \item[H2] \method gives lower welfare inequality than \baseline over all agents in a society.
\end{description}

\subsection{Simulation Setup}
\label{sec:sim-setup}
We modelled a sequence of CT games, described in Section~\ref{sec:environment}, between random pairs of agents in each multi-agent society. At each time step, all agents are randomly paired, and each pair of agents plays two rounds of CT. Each simulation was run over \timesteps time steps with a population size of \numagents to account for random variations in game setups and agent pairings at each time step. For each game round, we record the scores achieved by each agent. At the end of each simulation, we compute the metrics listed in Section~\ref{sec:metrics-hypotheses}. For \method agents, we initialise integral emotion to $E=0$, so that all agents start by using the policy associated with their SVO trait.

We present results from the average of three repeats of each simulation. We test for significant differences in our evaluation metrics between \method and \baseline, for whole societies and samples of agents with a particular SVO trait. We use a two sample t-test, and report the means, $\mu$, and p-values, $p$, and measure effect size as Cohen's $d$ \cite{Cohen-88-Statistics}.

\section{Results}
\label{sec:experiment-result}
To evaluate hypotheses H1 on collective welfare, and H2 on welfare inequality, we compared \method to \baseline for the four societies described in Section~\ref{sec:metrics-hypotheses}: \textit{altr-coop}, \textit{altr-self}, \textit{coop-self} and \textit{mixed}. 

Table~\ref{tab:society-metrics} compares population metrics measured for \method and \baseline agent societies: (1) mean game score (Mean Score) achieved by all agents in a society, and in samples of agents with the same SVO, as a measure of the collective welfare achieved by those groups; (2) the coefficient of variation (CoV) of the distribution of mean welfare for individual agents in each group as a measure of welfare inequality. All results are calculated from three repeat runs.  

\begin{table*}[!htb]
    \centering
    \caption{Comparing the mean score and coefficient of variation for samples from different agent societies with different SVOs.} 
    \label{tab:society-metrics}
    \begin{tabular}{llrSSSSSS}
    \toprule
        {Configuration} & ~ & ~ & {\baseline} & ~ & ~ & {\method} & ~ & ~ \\\cmidrule(r){1-3} \cmidrule(r){4-6} \cmidrule(r){7-9}
        {Society} & {Sample SVO} & {Size} & {Mean Score} & {Std} & {CoV} & {Mean Score} & {Std} & {CoV} \\ \midrule
         altr coop & all & 300 & 16.299 & 2.435 & 0.149 & 15.754 & 0.807 & 0.051 \\ 
        ~ &  altr & 150 & 13.877 & 0.206 & 0.015 & 14.966 & 0.169 & 0.011 \\ 
        ~ & coop & 150 & 18.720 & 0.221 & 0.012 & 16.541 & 0.177 & 0.011 \\\midrule 
         altr self & all & 300 & 15.257 & 7.358 & 0.481 & 15.456 & 1.676 & 0.108 \\ 
        ~ &  altr & 150 & 7.917 & 0.279 & 0.035 & 13.791 & 0.153 & 0.011 \\ 
        ~ & self & 150 & 22.597 & 0.271 & 0.012 & 17.121 & 0.181 & 0.011 \\ \midrule
        coop self & all & 300 & 16.299 & 1.758 & 0.108 & 15.816 & 0.690 & 0.044 \\ 
        ~ & coop & 150 & 14.558 & 0.226 & 0.015 & 15.147 & 0.157 & 0.010 \\ 
        ~ & self & 150 & 18.040 & 0.219 & 0.012 & 16.486 & 0.169 & 0.010 \\ \midrule
        mixed & all & 300 & 15.863 & 5.149 & 0.324 & 15.664 & 1.332 & 0.085 \\ 
        ~ &  altr & 100 & 9.269 & 0.271 & 0.029 & 14.016 & 0.170 & 0.012 \\ 
        ~ & coop & 100 & 16.527 & 0.243 & 0.015 & 15.731 & 0.160 & 0.010 \\ 
        ~ & self & 100 & 21.792 & 0.267 & 0.012 & 17.244 & 0.179 & 0.010 \\ \bottomrule
    \end{tabular}
\end{table*}

\begin{figure*}[!htb]
\centering

\begin{subfigure}{0.9\textwidth}
\includegraphics[width=1.0\columnwidth]{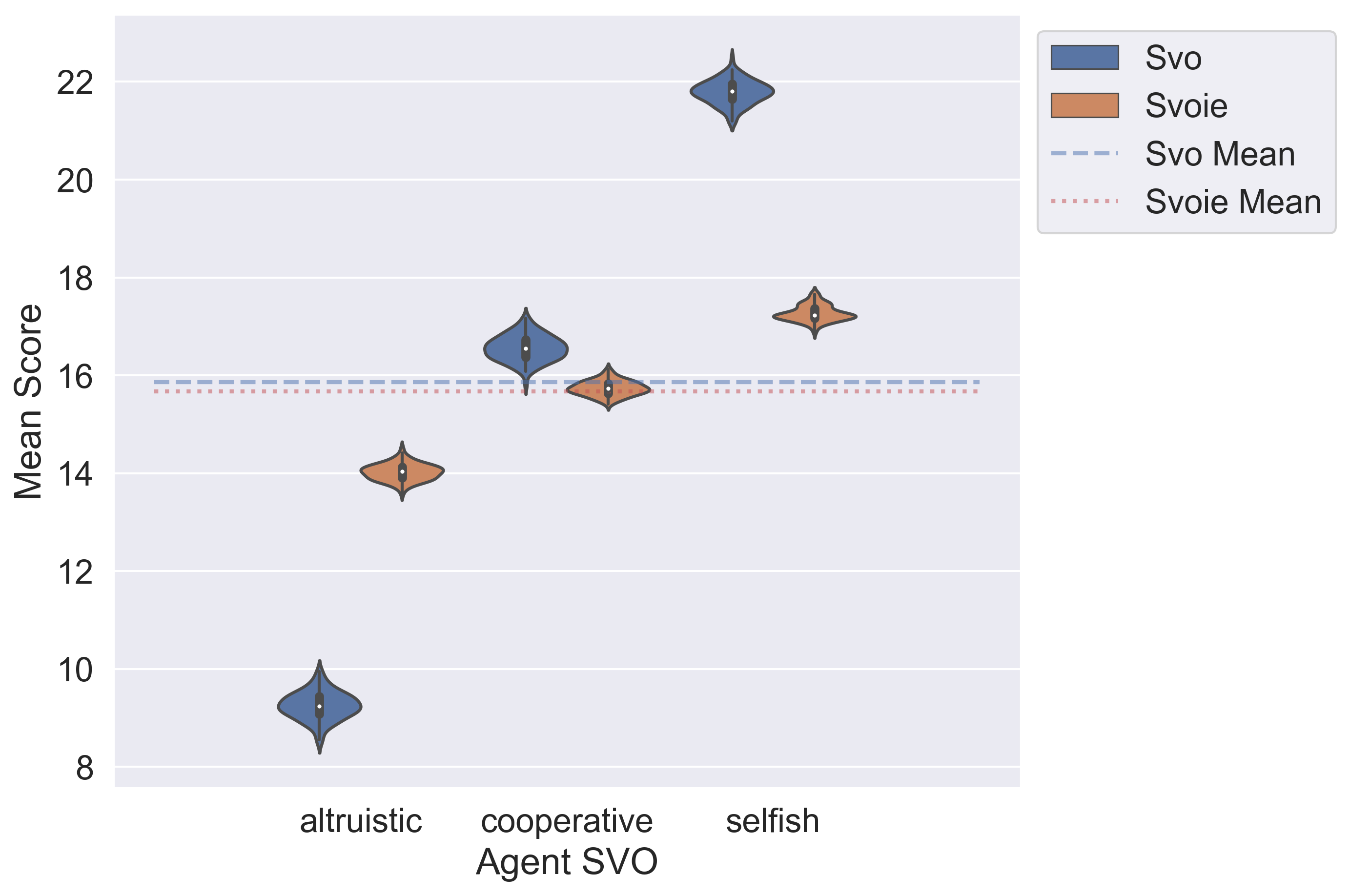}
\caption{Mean welfare.
}
\label{fig:score-violin}
\end{subfigure}

\begin{subfigure}{0.7\textwidth}
\includegraphics[width=1.0\columnwidth]{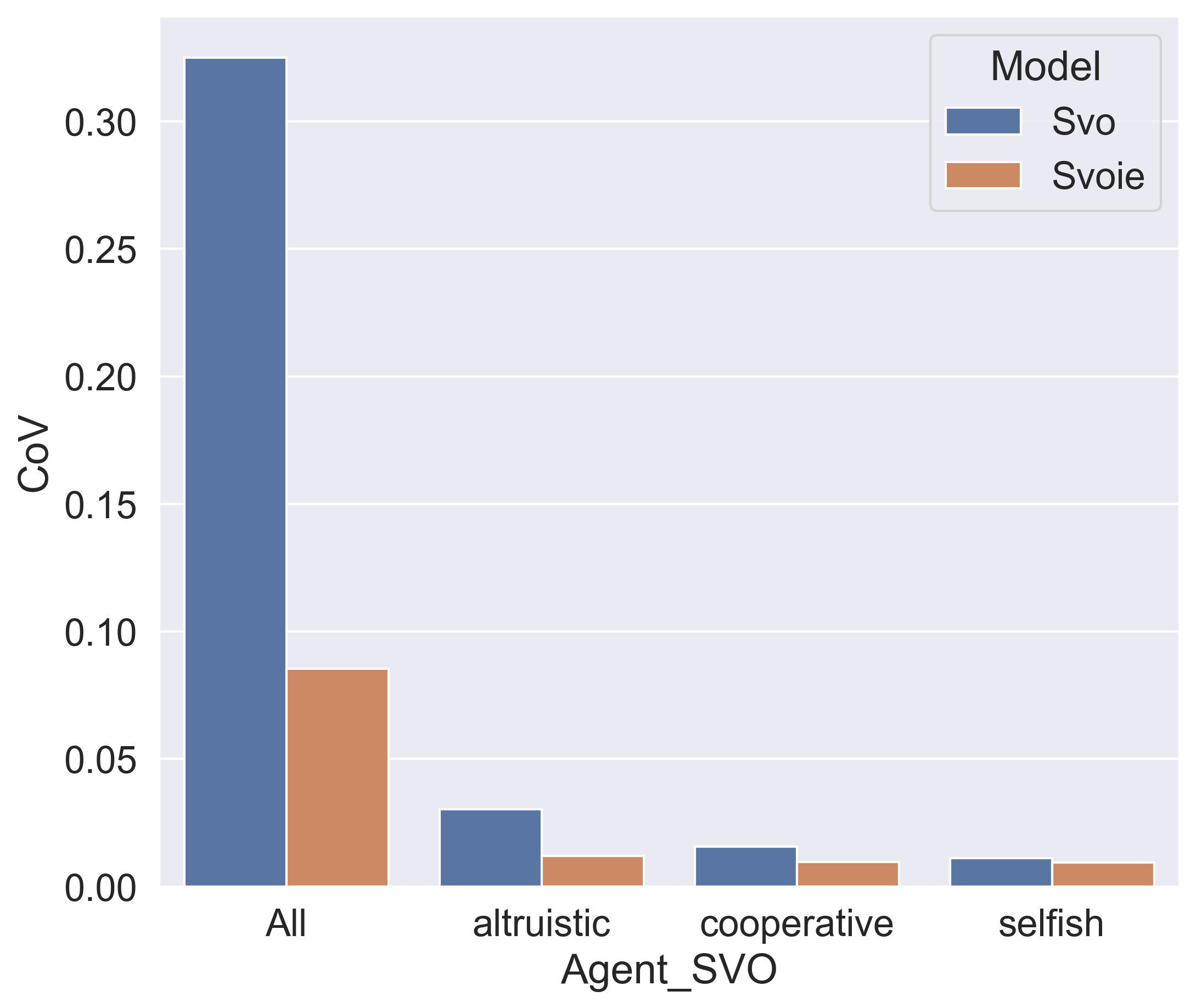}
\caption{Inequality.
}
\label{fig:cov-bars}
\end{subfigure}
\caption{Comparing welfare and inequality in societies of \baseline and \method with an equal mixture of altruistic, cooperative and selfish agents.}
\label{fig:comparision}
\end{figure*}

Our results indicate that deviating from stable SVO traits in \method has minimal impact on collective welfare. We find that there is no significant difference in collective welfare in the mixed society, \method ($\mu$ = 15.664) and \baseline ($\mu$ = 15.863), ($p$=0.5436, $d$=0.0497), or for the altr-self society, \method ($\mu$ = 15.456) and \baseline ($\mu$ =15.257)  ($p$=0.6739, $d$=0.034). However, \method yields lower collective welfare in both the altr-coop society, \method  ($\mu$ = 15.754) and \baseline ($\mu$ =16.299) (p$<$0.001, $d$=0.305), and in the the coop-self society, \method ($\mu$=15.816)  and \baseline ($\mu$ = 16.299) (p$<$0.001, $d$=0.371), albeit with small effect size. Therefore, the societies of \method agents, which are more likely to seek fair or ``inequity averse'' actions in response to reaching goals and which are more likely to seek unfair ``competitive equity averse'' in response to missing goals, were found to perform roughly as well as societies of agents which only act according to their SVO.  

We find that for all societies, \method agents yields significantly reduced welfare inequality compared to \baseline, with large effect size: (alt-coop: \baseline $\mu=$  16.299, \method $\mu=$ 15.754, p$<$0.001	$d$=84.106), (altr-self: \baseline $\mu=$ 15.257, \method $\mu=$ 15.456, p$<$0.001, $d$=39.543), (self-coop	\baseline $\mu=$ 16.299, \method $\mu=$ 15.816, p$<$0.001, $d$=73.007), (\baseline $\mu$ 15.863, \method $\mu=$ 15.664, p$<$0.001, $d$=32.384). This is illustrated by the distributions of individual welfare (mean score) for samples of agents in the mixed society simulation, shown in Figure~\ref{fig:score-violin} (a). We can see that the distributions of scores for each sample of agents with a particular SVO trait are further apart for the \baseline simulations and closer together for the \method simulations, but the ordering of their performance is unchanged. For example, we observe that altruistic \method agents perform better than altruistic \baseline agents, as they are likely to use unfair strategies in response to being taken advantage of, and selfish \method agents perform worse than selfish \baseline, as they are likely to use fair strategies after taking advantage of others. Further, the width of the distributions of mean score for each SVO is reduced in the \method simulation, therefore welfare inequality within an individual SVO trait sample is reduced relative to \baseline societies as well. This is reflected in the data shown in Table~\ref{tab:society-metrics} which contains measurements of the mean score achieved by samples of agents with different SVOs, and the coefficient of variation of the distributions of agent scores within those samples.

\section{Limitations, Directions and Conclusions}
\label{sec:conclusion}
We now discuss limitations and directions. Firstly, we model societies with heterogeneous SVO by generating populations of agents which can take one of either two or three different SVO traits, from altruistic, selfish and cooperative. In human societies, SVO varies continuously between individuals as described by the ring model \cite{liebrand_ring_1988}. Further, we assume an equal split of SVO traits in society, whereas in human societies, certain ranges of SVO are more common than others. Buckman \etal \cite{buckman_sharing_2019} implement a more realistic treatment of SVO in agent societies, by sampling agent traits from ranges of the SVO ring model found to be most prevalent in human society using relevant experimental data on SVO prevalence. We did not try to simulate realistic human societies, and were focused instead on modelling integral emotions alongside SVO to investigate how this would effect societal welfare and welfare inequality in a society of agents with different SVO policies. The three SVO policies we used in our work give different and non-overlapping distributions in welfare in our baseline simulations, and we therefore considered them to be appropriate for our purposes.      

Secondly, we model integral emotion as the variable state $E$ using several simplifying assumptions which prevent any direct comparison with integral emotion in real human behavior.  We only incorporate two emotion based policies, for positive and negative $E$ respectively. These policies are based on human behaviors which have previously been associated with positive and negative emotions, however they do not follow any explicit model. Further, we assume only one environment trigger, goal-achievement or non-achievement, to be relevant for influencing emotion, whereas there is evidence that other factors influence emotion, e.g. fairness of outcomes \cite{pillutla1996unfairness, straub1995experimental}, which could be utilised in the CT environment. We also only allow $E$ to vary over five possible states, and the extent to which $E$ changes is constant and chosen arbitrarily, preventing any differences in sensitivity to emotional stimuli between agents. We implemented \method agents as a course-grained model of SVO and integral emotion in agent societies, and did not seek to accurately model human behavior. In this context, we found that societies of \method agents had lower welfare inequality compared to baseline \baseline agents, and that collective welfare was preserved. These results suggest that by introducing transient changes in decision-making, triggered by task relevant events, agents can adapt their otherwise stable policies depending on the society they operate within. 

Agent-based modelling of social decision-making will always require simplifying approximations and assumptions, and cannot accurately capture all aspects of human behavior, but they are nevertheless useful for making predictions on specific aspects and edge-cases \cite{dignum_should_2023}. The limiting and simplifying assumptions of our agent model mean that we cannot predict whether the effects we observe would be found in a comparable study of human behavior. However, this simulation method offers a useful tool for modelling dynamic behavior, and better understanding existing theories of human behavior. Understanding the interplay between emotions and social preferences in human decision-making is important for the development of autonomous agents which can understand human social norms, and act in accordance with human moral and ethical principles \cite{Ajmeri-AAMAS20-Elessar,Kuipers2016HumanLikeMA,Murukannaiah-AAMAS20-BlueSky,Woodgate+Ajmeri-AAMAS22-BlueSky}. 
There is a rich body of existing work which explores human factors on norm emergence in multi-agent systems \cite{Morris+19:norm-emergence,savarimuthu_norm_2011}. 
\edit{Further, there are works which model how other factors may moderate rational behavior and individual preferences to explain seemingly irrational decisions made by humans, in games and other social contexts. For instance, Kampik \etal \cite{calvaresi_explaining_2019} investigated the role of sympathy in cooperative behavior, Sylwester \etal \cite{sylwester_homo_2013} have examined antisocial punishment as a form of social norm enforcement, and Köster \etal \cite{koster_spurious_2022} show how adopting arbitrary and inconsequential social norms may improve overall norm compliance in agent societies.} 

In our chosen simulation task environment, CT, random variations allow differences between the scenarios encountered by agents in each game, however the average performance for any agent is predictable over many time steps. This work could be extended by investigating how integral emotions influence societal outcomes across multiple task environments, to understand the implications of integral emotion for regulating behavior in a changing environment. Here, the societal effects of emotions and individual differences could be studied in the context of simulating the emergence and spread of norms in multi-agent systems which benefit survival. 

\section*{Acknowledgments}
DC was supported by the UK Research and Innovation (UKRI) Centre for Doctoral Training in Interactive Artificial Intelligence Award (EP/S022937/1). CH is a Leverhulme Research Fellow (RF-2021-533). NA thanks the University of Bristol for support.

\bibliographystyle{splncs04} 
\bibliography{references,Nirav,temp}

\end{document}